\documentclass[12pt]{iopart} 
\usepackage{graphicx}
\usepackage{epstopdf}
\begin{document}

\title[R. Debbe QM2008]{Recent results from BRAHMS}

\author{R. Debbe for the BRAHMS Collaboration}
\address{ Physics Department Brookhaven National Laboratory, Upton, NY, 11973, USA}
\begin{abstract}

The BRAHMS collaboration ended its data collection program in 2006. We are now well
advanced in the analysis of a comprehensive set of data that spans systems ranging in
mass from p+p to Au+Au and in energy from $\sqrt{s_{NN}} = 62.4$ to 200 GeV. Our
analysis has taken two distinct paths: we explore the rapidity dependence of intermediate
and high-transverse-momentum, identified-particle production, thus helping to
characterize the strongly-interacting quark-gluon plasma (sQGP) formed at RHIC; we
also explore particle yields at lower transverse momentum to develop a systematic
understanding of bulk particle production at RHIC energies.

\end{abstract}

\section{Introduction} The time evolution of the hot and dense medium formed at RHIC appears to be well described in
terms of the hydrodynamic expansion of a low-viscosity fluid followed by a hadronic
cascade \cite{Shuryak}. Different aspects of the BRAHMS data can be used to help elucidate the
expansion and subsequent hadronization behavior. Comparisons of intermediate and
high transverse momentum ($p_{T}$) distributions from A+A systems to corresponding, properly scaled p+p results
over a wide range in rapidity contribute to the characterization of the sQGP by
highlighting effects that cannot be attributed to the underlying nucleon-nucleon
collisions. The study of bulk particle and strangeness production helps to develop the
thermodynamical aspects of the reaction. Studying the net proton production as a
function of rapidity reveals aspects of the baryon transport in the reaction. This contribution
describes recent results from the BRAHMS collaboration.
\section{Energy loss and suppression at high rapidity}  To help frame the
discussion, 3+1D hydrodynamic calculations of T. Hirano  \cite{HiranoNara} are used as a guide for
understanding some of the observed intermediate- and high-$p_{T}$ results. This model 
assumes 
 the establishment of a partonic medium that has reached local thermal equilibrium in a particularly short time (~0.6 fm/c), and starts with a maximum energy density of 
 $34.2\ \mbox{ GeV}/fm^{3}$ for Au+Au collisons at 200 GeV.  The bulk properties of that medium are later reflected in  the particles produced with $p_{T}<1.5\ \mbox{GeV}/c$. Partons generated in hard  $2 \rightarrow 2$ QCD 
interactions obtained from \textsc{PYTHIA} \cite{Pythia} are also part of the 
calculation to account for the higher $p_{T}$ end of the spectra.

\begin{figure}[!ht]
\begin{center}
\resizebox{1.0\textwidth}{!}
{\includegraphics{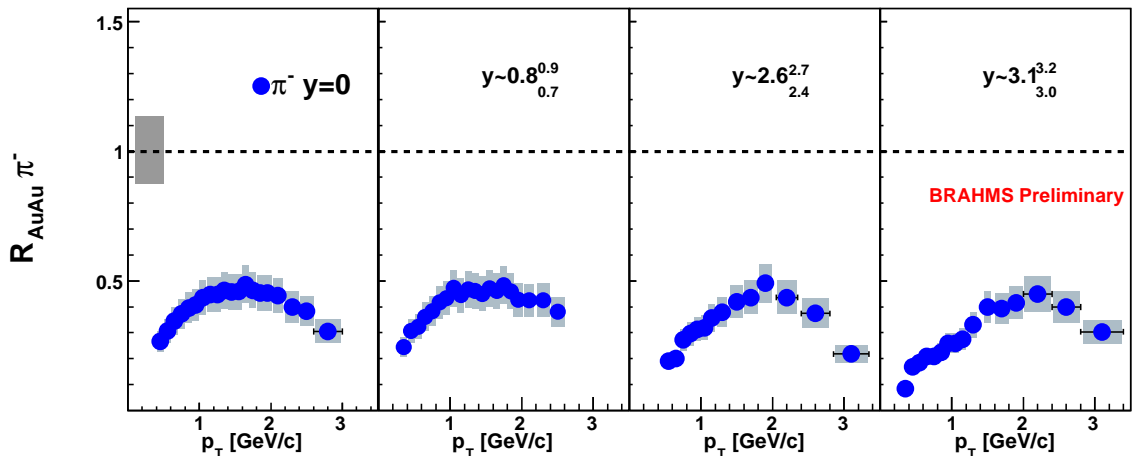}}
\end{center}
\caption{\label{fig:Raa200} The NMF for negative pion production measured in central ($0-10\%$) Au+Au collisions at 200 GeV. 
Statistical errors are smaller than the symbols and the estimated systematic errors are displayed as shaded boxes. The estimated error on the number of binary collisions 
is shown with the gray box centered at 1.} 
\end{figure}

Nuclear effects in A+A collisions can be explored using the Nuclear Modification Factor (NMF).  This $p_{T}$ dependent ratio compares the yield of moderate and high $p_{T}$ particles detected from A+A system to the ones produced in  p+p collisions (at RHIC the measurements can be done at the same energy using the same apparatus)  normalized to the Glauber number of binary collisions in the A+A system. A value of this factor close to one is construed as the absence of nuclear effects as the A+A collisions are seen as an incoherent sum of p+p interactions.  This factor was first used in p+A studies where it surpassed unity at intermediate values of $p_{T}$ in what is now understood as multiple interactions in the A target.
The NMF  can also be smaller than one indicating some sort of shadowing, a deficit of scattering points in the initial state, or alternatively reflecting energy loss at
partonic level.
The NMF  has been used to describe earlier Au+Au  results  at RHIC in terms of this energy loss \cite{PRL91}. 
In  d+Au collisions a clear enhancement at intermediate $p_{T} $ values around mid-rapidity changing into a suppression at high rapidity \cite{dAPRL} has been seen as a possible indication of the onset of gluon saturation. The NMFs from central Au+Au collisions at 200 GeV measured by BRAHMS 
show a remarkable feature; the factor is practically unchanged as function of  rapidity 
from $y = 0\ \mbox{to}\ y=3.1$, see Fig. \ref{fig:Raa200}.

The hydrodynamical model is used to guide our understanding of the observed weak rapidity dependence of the  NMFs obtained from our most central events.  The transit of high 
energy partons through the expanding medium can be followed starting from their 
generation point.
At each step of the hydro calculation the energy of the parton and the local density of 
the medium are used to calculate the fractional energy loss by gluon emission using the Gyulassy, Levai and Vitev model \cite{GLV}. After approximately 8 fm/c the system has expanded and cooled to a level where  the interactions of the partons with the medium end. The initial and final values of the energy density at 
each rapidity are listed in Table 
\ref{tab:hydroTable200}. The average total fractional energy lost by partons 
is also listed in this table and it shows a drop by almost a factor of 2 as rapidity 
changes from 0 to 3.
\begin{table}
\begin{center}
\begin{tabular} {|l|c|c|c|c|}
\hline
Rapidity & y = 0
 & y = 1
 & y = 2
 & y = 3 \\
 \hline
$\epsilon(0,0)$ [$GeV/fm^3$] at  $\tau = 0.6 fm/c$ & 34.2 & 34.2 & 34.2 & 21.2 \\
$\epsilon(0,0)$ [$GeV/fm^3$] at  $\tau = 8.4 fm/c$ & 5.3 & 5.2 & 4.6 & 2.8 \\
Average fractional energy loss   & 17.2 \% & 16.9 \% & 15.2 \% & 10.1 \% \\

  \hline
\end{tabular}
\caption{Initial and final energy density at the center of the system  at four rapidity values, as well as average energy loss for central Au+Au hydro simulations.} 
\label{tab:hydroTable200}
\end{center}
\end{table}

 
If one were to naively translate the factor of two drop in energy loss shown  in Table 
\ref{tab:hydroTable200}
into NMF's, one would see a gradual increase from its value at mid-rapidity to values 
closer to one. However, it should also be noted that spectral shapes do change with 
rapidity. The increased steepness of the cross sections at high rapidity, resulting 
from  phase space limitations, might offset 
the  reduced energy loss there, keeping the observed suppression constant.
 Other 
effects like shadowing or the onset of saturation in the nuclei wave functions at RHIC 
energies may be contributing factors to the results shown in Fig. \ref{fig:Raa200}. The 
data calls for a detailed theoretical study before firm conclusions can be reached.

\begin{figure}[!ht]
\begin{center}
\resizebox{1.0\textwidth}{!}
{\includegraphics{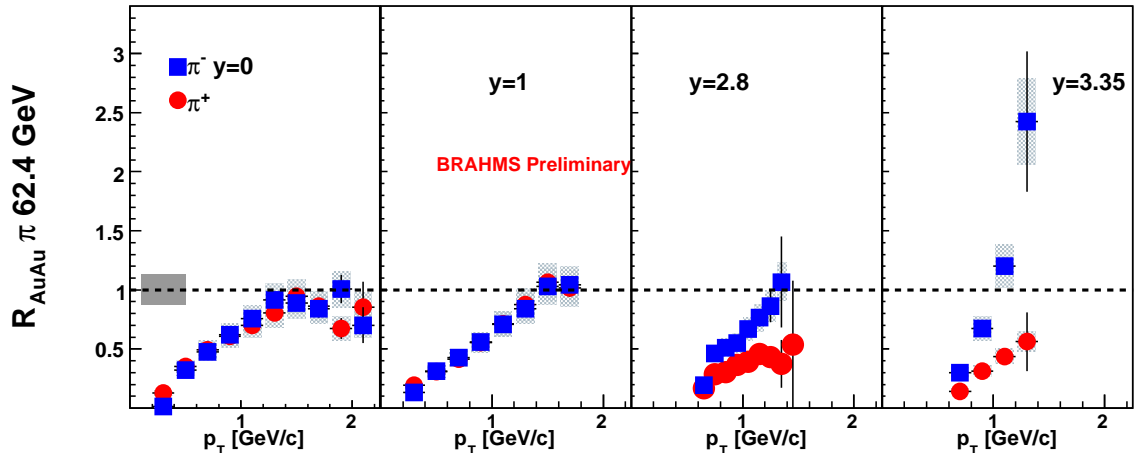}}
\end{center}
\caption{\label{fig:Raa62} Nuclear Modification factor extracted at four rapidity values for positive and negative pions detected in 0-10\% central Au+Au collisions at 62.4 GeV. Positive pions are shown with filled circles (red online) and negative pions are shown with filled squares (blue online). Statistical errors are shown with vertical lines at each point, and systematic errors are shown with shaded boxes. }
\end{figure}

We have also collected and analyzed data at lower energy (62.4 GeV) in p+p and A+A systems.  We report the following results also  within the context of a fluid expanding hydrodynamically. 
Figure \ref{fig:Raa62} shows the NMFs for positive and negative pions at four rapidity values. The
 first striking difference when comparing Fig.  \ref{fig:Raa62} to Fig \ref{fig:Raa200} is the completely 
different rapidity dependence at the two energies.  Around mid-rapidity (up to $y \sim 1$) the NMF's for positive and negative pions are equal, as if isospin effects are canceled,
 as was also the case at high energy, but now there is no evidence of a suppression at higher $p_{T}$. At forward rapidities the NMF of negative and positive pions are different and their difference grows more pronounced with $p_{T}$. This difference is 
related to the isospin of the p+p system. 
According to the hydro calculation at this energy, there is small energy loss at $y=0\ \mbox{and}\ y=1$ 
and practically no energy loss at higher rapidities, well in agreement with what is shown if Fig. \ref{fig:Raa62}.
 More details about BRAHMS high $p_{T}$ analysis can be found in these proceedings \cite{Ian}. 

\section{Bulk particle production} The BRAHMS Collaboration has measured identified particle production in p+p and Au+Au at $\sqrt{s_{NN}} = 62$ and 200 GeV. In this section we describe the pion production density in rapidity space, $dn/dy$. Invariant yields  for 
 identified particles are constructed from sets of particles detected in the two BRAHMS 
spectrometers at different angles in narrow rapidity windows. These invariant yields are fully corrected for  geometrical acceptance, tracking inefficiencies, and particle identification efficiency as well as losses resulting from
 multiple scattering, decay in flight and nuclear interactions along the paths in the 
spectrometers. 

\begin{figure}[!ht]
\begin{center}
\resizebox{1.0\textwidth}{!}
{\includegraphics{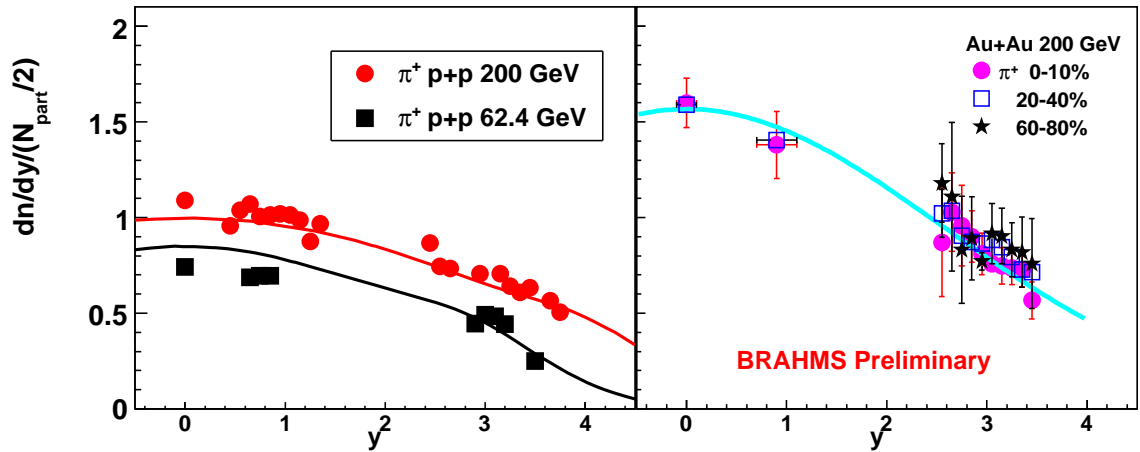}}
\end{center}
\caption{\label{fig:dndy} Positive pion densities in rapidity space scaled by the number of participant nucleon pairs. The left panel shows the pions produced in p+p collisions at 200 GeV (normalized to total inelastic cross section of 41 mb) with filled circles (red online) and at 62.4 GeV with black symbols. The smooth curves show the corresponding density obtained with \textsc{PYTHIA}. The panel on the right  shows positive pions produced in Au+Au at 200 GeV in 5 centrality bins. For clarity, the overall contribution of the spectra 
statistical and systematical errors as well as the fit uncertainties are displayed for 
0-10\% and 60-80\% centrality samples. The smooth curve is a single Gaussian fit to the most central sample.  }
\end{figure}

We present the yields extracted in p+p collision at 200 and 62.4 GeV as well as the centrality-dependent particle densities in Au+Au collisions at 200 GeV where we investigate 
how the size of the colliding system's overlap region affects the production of 
particles. 
Total yields are found by integrating over  $p_{T}$. Because our measurements do not 
extend to low values of $p_{T}$ an extrapolation is necessary to evaluate the integral. 
Several functions were used to fit the $p_{T}$ distributions. The pion distributions from 
p+p collisions at 200 GeV were
 fitted with the Levy function, the pions at 62.4 GeV are well described by power law shapes around mid-rapidity and by single exponential functions at high rapidity.  Finally so
called ``blast-wave'' fits were used to obtain   the pion and proton yields in Au+Au   
collisions at 200 GeV shown in the right panel of Fig. \ref{fig:dndy}.  The hydrodynamically based blast wave fits included the  spectra of pions and protons and their anti-particles at different rapidity values. The velocity profile as a function of the system
 radius was extracted at mid-rapidity and  later fixed for the fits at higher rapidity. The choice of this functional form is motivated by the existence of strong radial flow in Au+Au collisions and the very low $p_{T}$ points measured by PHOBOS \cite{lowPtPHOBOS} which exclude functions with strong rise as $p_{T} \rightarrow 0$. 
Details about similar analyses performed in p+p and d+Au collisions can be found in these proceedings \cite{Hongyang}.

The integrated yields are shown in Fig. \ref{fig:dndy} where we display rapidity densities for positive pions produced in p+p collisions at $\sqrt{s_{NN}}$ = 200 and 62.4 GeV 
in the left panel. The right panel  shows the same  rapidity densities for pions 
produced in Au+Au at 200 GeV in three centrality samples normalized to the number of 
participant nucleon pairs $N_{part}/2$ obtained with a Glauber Monte-Carlo calculation. 
Within the errors quoted for all of 
these measurements one interesting feature appears at first glance, all these distributions can be described with single Gaussian shapes. Furthermore, the Gaussian  widths 
follow closely those expected from the Landau-Carruthers hydrodynamical model 
\cite{Carruthers} even though that model makes use of the same equation of state (EOS) 
with a speed of sound fixed at $c_{s}^{2} = 1/3$ at all energies and does not include 
a phase transition. The significance of this agreement is not clear since more 
sophisticated 1+1D models which include phase transitions and different EOS have been 
used to describe successfully  the BRAHMS results in central Au+Au collisions at 200 GeV 
and stand in clear conflict with the simple Landau model \cite{Satarov}.   It can also be 
seen in Fig. \ref{fig:dndy} that the particle density per participating nucleon pair at 
mid-rapidity are considerably greater in the 
A+A system as compared to the p+p system at the same energy. These scaled densities 
remain almost 
constant for the different centrality samples of the high energy Au+Au collisions. 
 
\section{Strangeness production as function of $\mu_{B}$} 
\begin{figure}[!ht]
\begin{center}
\resizebox{1.0\textwidth}{!}
{\includegraphics{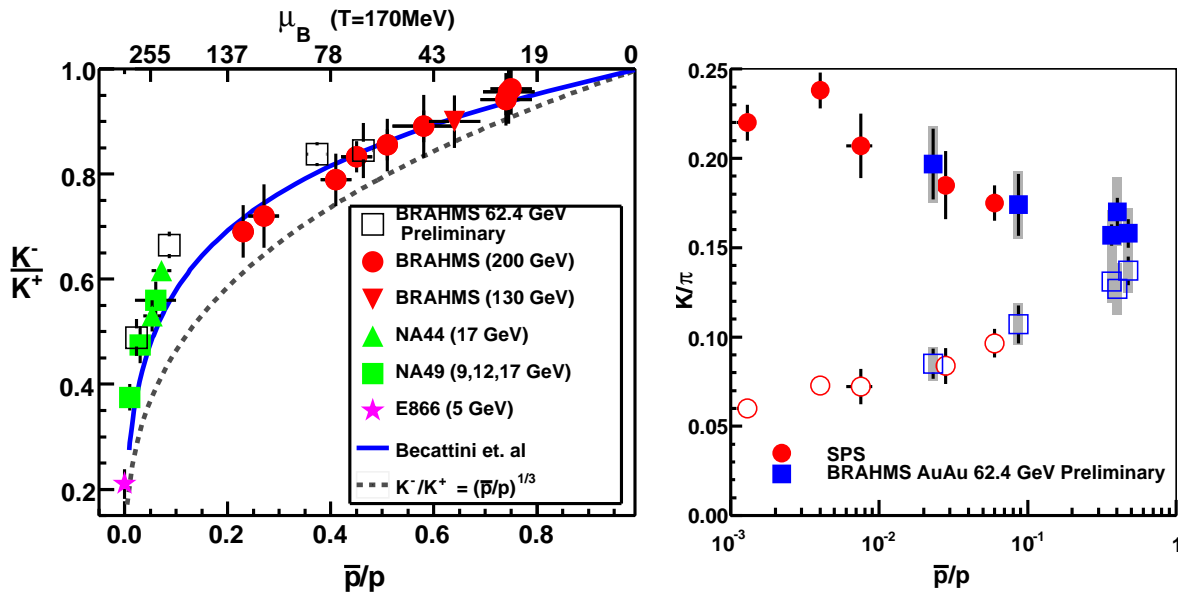}}
\end{center}
\caption{\label{fig:KoverPi} (Left) Anti-particle over particle ratios for kaons and protons in different colliding A+A systems together with a curve based on a statistical model of hadronization.  (right) The kaon over pion ratios as function of the anti-proton to proton ratio at the SPS is shown with opened and filled circles (red online) for negative and positive particles respectively. The BRAHMS points extracted from high rapidity Au+Au collisions at 62.4 GeV are shown with square symbols (blue online, filled: positive, open: negative). Systematic errors are shown with gray boxes. }
\end{figure}

A very interesting correlation
 is shown in the left panel of Fig. \ref{fig:KoverPi} : The ratios of anti-particle to 
particle for kaons and protons yields measured in different A+A systems at energies 
ranging from AGS ($\sim 5\ \mbox{GeV}$), SPS (9 to 17 GeV) and RHIC lie on a common curve. 
Moreover, the RHIC points at different rapidities also span the same curve. This correlation is well described by statistical models of hadronization \cite{Becattini} and 
\cite{Resonance} that call for local chemical equilibrium at a common temperature of 170 MeV and a baryochemical potential $\mu_{B}$ in a one to one relation with the anti-proton to proton ratio. Within the same experiment, the  value of  the baryochemical potential at the late stages of the collisions can be increased if the measurements are made at higher rapidites. In particular, we have studied kaon and proton yields in Au+Au collisions at 62.4 GeV where the values of $\mu_{B}$ at the highest measured rapidities  are expected to be in the range of those extracted at SPS. 

The right panel of Fig. \ref{fig:KoverPi} shows the ratio of kaon to pion yields, a 
measure of strangeness production, versus the corresponding anti-proton to proton ratios. The mid-rapidity SPS points shown with open and filled circles (red online) display the high energy side of the so called "Marek Horn"  and the BRAHMS results at high rapidity
 displayed with open 
and filled squares (blue online) show strikingly similar behavior to the SPS 
mid-rapidity measurements at their highest energy \cite{NA49} in what may be considered 
as an indication of their chemical equivalence. More details on this analysis can be found in these proceedings \cite{Ionut}.
  
\section{Baryon transport} 
\begin{figure}[!ht]
\begin{center}
\resizebox{1.0\textwidth}{!}
{\includegraphics{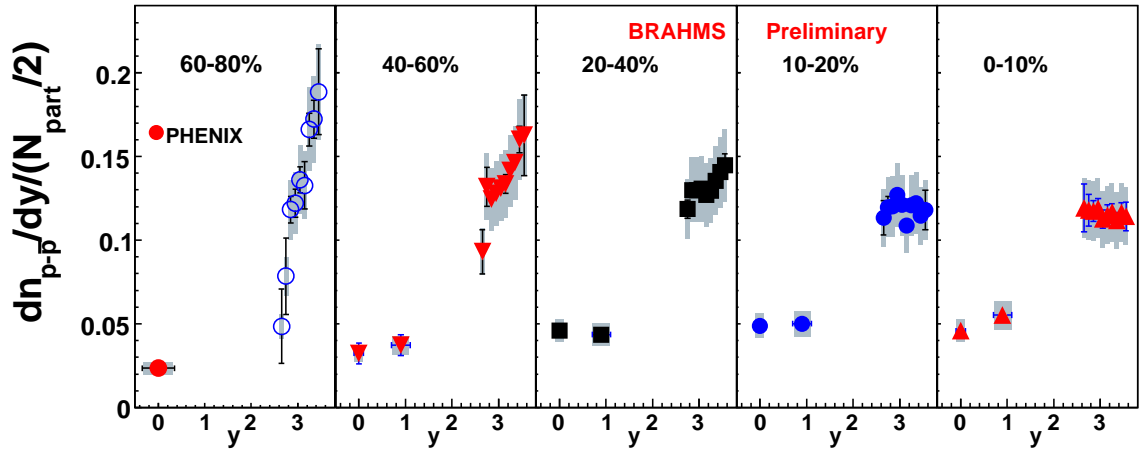}}
\end{center}
\caption{\label{fig:netProton}Net proton yields in Au+Au collisions at 200 GeV normalized 
by $N_{part}/2$ for different centrality samples. The $y\sim0$ and $y\sim1$ points at all 
centralities were extracted from a previous BRAHMS publication \cite{PRC72} 
 except the single PHENIX point  \cite{PHENIXPRC69} in the  60-80\% centrality sample. No feed down corrections are applied. }
\end{figure}

Baryon number is a conserved quantity. Before the collision, 
all baryons are localized at beam rapidity, but once the collision occurs the total baryon number of the colliding system spreads across rapidity space. The way that rearrangement happens can give great insight into the details of the nuclei-nuclei interaction. The BRAHMS experiment at RHIC, with its ability to cleanly identify  protons has  collected a 
large set of data that is now being used to study baryon transport in simple systems like 
p+p collisions at two energies as well as A+A collisions. By selecting the centrality of 
the 
A+A collision one can control the size of the interaction region and extract information
 on how baryon number transport depends on this geometry.
The rapidity distribution of net-proton ($p-\bar{p}$) is used to study how the beam protons are transported in rapidity space. The net-proton distributions, scaled by the number of 
participating nucleon pairs corresponding to five centrality samples, are displayed in 
Fig. \ref{fig:netProton}.  Our most peripheral events (60-80\%) together with an 
additional PHENIX point measured at y=0 are shown in the left-most panel of the figure 
and panels show increasing centrality towards the 0-10\% on the right. The values of 
net-proton yields have not been corrected for feed-down contributions from weak decays, and the values  extracted from previous publications have been scaled back to their uncorrected values. The scaled net-proton distribution for the most peripheral events are
similar to the ones extracted from p+p at the same energy; the distribution is strongly peaked in the 3-4 rapidity range. As the volume of the colliding system  increases, one can see a clear 
rearrangement   with more protons being shifted all the way to mid-rapidity as the high rapidity peak seen in peripheral events is being transformed into much more rounded shape. This evolution of the net-proton yields with the size of the colliding system, together with a forthcoming distribution extracted from p+p collisions as well as low energy A+A systems should constitute a data sample that will constrain all baryon transport models.

\section{Summary} We have contributed to the characterization of the sQGP with our ability to compare the 
yields of identified particles at intermediate and high $p_{T}$ to appropriately scaled 
yields of the same particles measured in p+p collisions with the same apparatus. 
At the highest energy our studies of the NMFs of pions at four different rapidity 
values, together with insights obtained from 3+1D hydrodynamical model lead us to 
consider the possible presence of phenomena that compensate for a reduced effect of 
energy loss.
We also report on our ongoing work to characterize the systematics of bulk particle 
production. In particular,  we described pion  densities in p+p and Au+Au at 200 and 62.4 GeV. All of these distribution display a simple Gaussian shape and appear to scale with the number of participant nucleons.
As an aside to our work on bulk particle production we described work done in Au+Au at 62.4 GeV to study strangeness production at high rapidity and how it reproduces high energy SPS results. This confirms the fact that in the late stages of the interaction, hadronization proceeds according to  an statistical distribution driven mainly by the local value of the baryochemical potential.    
Finally, we  report on our work to establish the baryon transport in Au+Au collisions 
at 200 GeV  studying the distribution of net-protons in rapidity space for samples of events with different centralities.
By changing the centrality of the events and with the help of existing transport models,
we expect to be able to identify the effect of the interaction volume size on the rates of absorption or re-interaction, as reflected 
in the yields of protons and anti-protons.


\section*{References}
\begin{thebibliography}{10}
\bibitem{Shuryak}E. Shuryak, these proceedings and references therein.
\bibitem{HiranoNara} T.Hirano and Y.Nara, Nucl. Phys. {\bf A}743, (2004) 305.
\bibitem{PRL91}I. Arsene {\it et al.}  Phys. Rev. Lett. 91, 072305 (2003).
\bibitem{dAPRL}I. Arsene {\it et al.} Phys. Rev. Lett. 93, 242303 (2004).
\bibitem{Pythia} T. Sj¬ostrand et al., Comp. Phys. Comm. 135, 238 (2001).
\bibitem{GLV}M. Gyulassy, P. L«evai, and I. Vitev, Nucl. Phys. {\bf B}594,371 (2001).
\bibitem{Ian} I. Bearden, these proceedings.
\bibitem{Carruthers} P. Carruthers and M. Duong-van, Phys. Rev. {\bf D}8, 859 (1973).
\bibitem{Hongyang} H. Yang, these proceedings.
\bibitem{lowPtPHOBOS} B. B. Back {\it et al.} Phys. Rev. {\bf C}70, 051901(R) (2004).
\bibitem{Satarov} L.M.~Satarov, A.V.~Merdeev, I.N.~Mishustin and H.~Stoecker,
  Phys.\ Rev.\  C {\bf 75}, 024903 (2007).
\bibitem{Becattini} F. Becattini, J. Cleymans, A. Keranen, E. Suhonen, and K.
Redlich, Phys. Rev. C 64, 024901 (2001).
\bibitem{Resonance}B. Biedro$\acute{n}$ and W. Broniowski Phys. Rev. {\bf C}75, 054905 (2007).
\bibitem{NA49} NA49 Collaboration, Phys.Rev. {\bf C}73, 044910 (2006).
\bibitem{Ionut} I.C. Arsene, these proceedings.
\bibitem{PRC72}I. Arsene {\it et al.}  Phys. Rev. {\bf C}72, 014908 (2005).
\bibitem{PHENIXPRC69} S.S. Adler  {\it et al.}  Phys. Rev. {\bf C}69, 034909 (2004).

\end{thebibliography}
\end{document}